\documentclass{camera}
\usepackage{graphicx}  

\begin{document}

%
\title{The nuclear matter phase diagram, multifragmentation, Zipf's law}

%
\author{Wolfgang Bauer, Brandon Alleman \and Scott Pratt}

%
\organization{Department of Physics and Astronomy\\
Michigan State University\\
East Lansing, MI 48824-2320, USA\\
URL: http://www.pa.msu.edu/$\sim$bauer/\\
email: bauer@pa.msu.edu}

\maketitle

\begin{abstract}
Some of the progress in determining the phase boundaries of the nuclear phase diagram, the location of the critical point of the nuclear fragmentation phase transition, and the values of the critical exponents of this transition is reviewed. The recently postulated connection of the size of the largest fragments as a function of their rank with Zipf's law known from linguistics is discussed, and an exact formula is derived. We show that Zipf's law is not strictly observed in nuclear fragmentation, but that one can derive useful approximations. 
\end{abstract}

%

\begin{figure}[hbt]
\centerline{\includegraphics{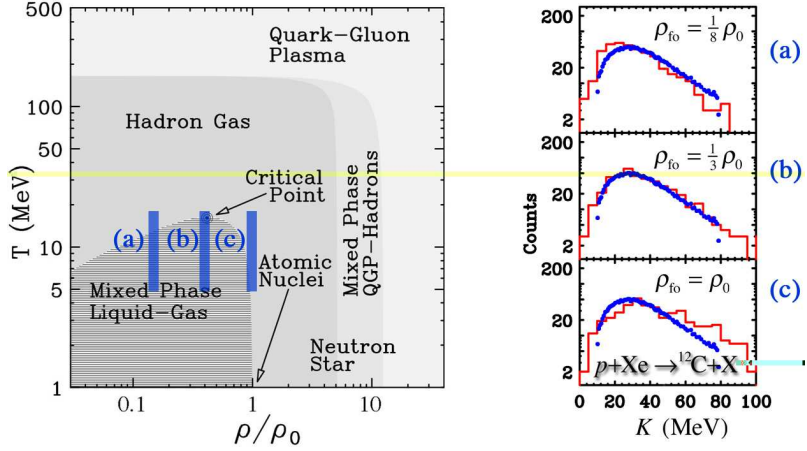}}
\caption{Left: General features of the nuclear matter phase diagram. Right: circles: experimental data
for the energy spectra of $^{12}$C fragments produced in 300 GeV proton-induced fragmentation
events of Xe \cite{hir84}; histograms: calculations with different values of the freeze-out density.
The three cases on the right correspond to the areas of the phase diagram indicated on the left.}
\label{fig1} 
\end{figure}

One of the premier goals of nuclear physics during the last two decades has been the determination
of the phase diagram of nuclear matter and finite nuclei.  From basic many-body physics, from
symmetry arguments, from general considerations of the nucleon-nucleon force
and of QCD, as well as from the analysis of experimental data, we have a fairly well established idea
what the gross features of this phase diagram are.  There are two distinct phase transitions, one
between hadrons and quark/gluons, and the other between a nuclear ``liquid'' and a vapor of hadrons,
see fig.~\ref{fig1}.

In order to learn about the parameters of the nuclear equation of state, we have to overcome several
severe challenges.  First and foremost, we cannot prepare our system at a given value of the state
variables and keep it there for an extended time interval.  The only way in which we can excite our
nuclear system is by colliding finite nuclei with each other or with small probes such as individual
protons or pions. 

A lot of effort has been devoted to constructing ways in which to provide reliable and necessarily
indirect measurements of the state variables. 
A priori, it is an open question whether thermodynamic equilibrium is achieved,
and if equilibrium is achieved, at what time during the course of the reaction
this happens first, and at what time different degrees of freedom freeze out.
Further, it is unclear on which
path the excited nuclear system migrates through the above phase diagram.

However, the last few years of investigation have given rise to the hope
that we can use nuclear reactions at certain energies, impact parameters,
and target/projectile combinations to enable the excited nuclear system
to pass sufficiently close to the critical point and stay there long enough
for the final fragmentation pattern to contain evidence on the location of
the critical point in the $\rho/T$ diagram and perhaps the universality class of the
phase transition.

But how can a system with numbers of constituents of only several hundred at most can enable us to
make reliable statements about a phase transition that is defined strictly only in the infinite size limit?
We would like to argue that nuclear physics has a privileged position 
and can provide experimental and theoretical answers to this question that are relevant for
many-body science in general.

In order to probe the fragmentation phase transition one can either use symmetric heavy ion collisions
at beam energies per nucleon around the Fermi energy or by using protons or pions as projectiles
on heavy targets in the limiting fragmentation regime ($E_{\rm lab} > 10$ GeV).  Of course, reverse
kinematics is also possible in the latter case, as is the use of light ions with a corresponding energy
deposition.  While the use of light ions already leads to cavitation \cite{wang96}, the use of near symmetric projectile-target combinations leads to a significantly higher compression that is released
into a high value of radial flow, which can cause the appearance of bubble-nuclei \cite{bbs92}. While
the transient creation of bubble nuclei is of course interesting in its own right, it can also lead to a
significant change in the observed fragment spectra \cite{pbc93}.

We have introduced the percolation model of nuclear fragmentation already two decades ago
\cite{bau84,bau85,bau86,bau88}. It represents the nucleons by a collection of lattice sites on a three-dimensional lattice (usually simple cubic, but not necessarily) and their interactions by 
nearest-neighbor bonds. One can show \cite{bau88} that
this prescription approximately reproduces the Bethe-Weizs\"acker binding energy systematics. Energy is deposited via the random breaking of bonds, and sites still connected to each other are identified as
fragments. The number of broken bonds is strictly proportional to the total energy deposited; this
calculation can be done in a microcanonical way (fixed number of bonds), a a canonical way, 
relating the bond breaking probability to the temperature via \cite{Li93}
$p_{\rm b} = 1 - 2\Gamma\left(3/2,0,B/T\right)/\sqrt{\pi}$.
We have used the Glauber approximation \cite{bau86}
or intranuclear cascades \cite{bot95} or experimental data \cite{kbb02} to estimate the energy
deposition and residue size after pre-equilibrium emission as a function of impact parameter.

The calculations on the right hand side of fig~\ref{fig1} (histograms) were performed with this
percolation model and a subsequent multi-body Coulomb expansion. The observed fragment
distributions are sensitive to the freeze-out density and thus enable us to determine
this important parameter. The calculated fragment spectra are of Boltzmann-shape, but the
slope of the spectra is {\em not} equal to the intrinsic temperature of the hot nuclear system at
breakup \cite{bau95}.

\begin{figure}[hbt]
\centerline{\includegraphics{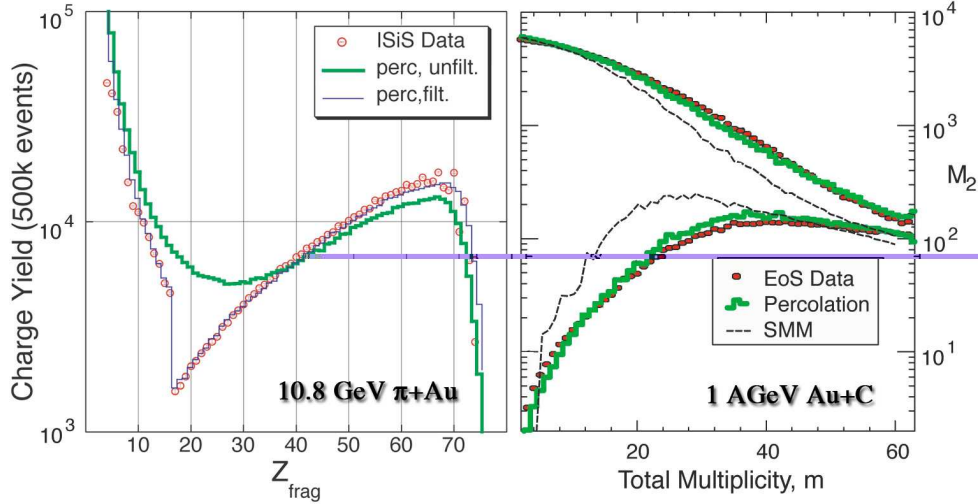}}
\caption{Left: Charge yields from the reaction of high energy pions with gold, data from \cite{kbb02};
right: event-by-event distribution of the second moment of the fragment charge
distribution as a function of the multiplicity
including (upper curves) and excluding (lower curves) the largest fragment in each event, data from
\cite{Rit95}.}
\label{fig2} 
\end{figure}

As a further sampling of the successes of the percolation model we 
show in fig.~\ref{fig2} on the left a
comparison of the charge yields in the reaction of 10.8 GeV pions with gold nuclei \cite{kbb02}
and on the right an event-by-event reconstruction of the second moment of the
fragment charge distribution as a function of multiplicity \cite{Gil94,Rit95}
for the inverse kinematics reactions of 1 A\,GeV
gold nuclei on a carbon target.

The percolation model has a well-defined universality class in the infinite size limit. The cluster size distribution in the vicinity of the critical point is governed by the critical exponents $\sigma$ and $\tau$ 
and follows the scaling law
\begin{equation}\label{eqa}
n(A,p) = a\,A^{-\tau}f(A^\sigma\cdot(p-p_c)) \quad \quad
  (\textrm{for } p \approx p_c)
\end{equation}
where $A$ is the cluster size, $p$ is the control parameter and $p_c$ its critical value. $f$ is the scaling
function with the property $f(0)=1$. Thus, if one plots the log of the scaled yield, $n_s(p) s^{\tau}$,
versus the scaled control parameter, $(p-p_c) s^\sigma$, the cluster distributions produced in
percolation calculations all fall on a straight line that crosses the point (0,1).

One can also conduct an analysis of experimental fragment size distributions, using the value of the
critical temperature and those of the critical exponents $\sigma$ and $\tau$ as free parameters, then
performing a $\chi^2$ optimization procedure to find the set of the values for which the scaling
collapse of the data is optimized. This was done in \cite{kbb02} for the 10.8 GeV pion-induced gold
fragmentation data. The values of the critical parameters obtained
are $\sigma=0.5 \pm 0.1$, $\tau = 2.35 \pm 0.05$, and $T_c=8.3 \pm
0.2$ MeV. However, we need to point out that this analysis is not entirely model independent,
because one needs to correct for the effects of sequential decays and feedings.  Similar values 
for the critical exponents were reported in \cite{ell02}.


Recently, Zipf's Law has been applied to the size of the largest cluster in nuclear fragmentation.
Zipf's Law in its original formulation applies to the frequency of the most frequent words in the
English language. Zipf found \cite{zipf} in the 1940s
that if one ranks the words by their frequency, one finds
that the frequency $F$ has a power-law dependence on the rank $r$: $F(r) = c\,r^{-\lambda}$,
with $c$ = constant, $\lambda\approx 1$). The same distribution was found much earlier by Pareto in ranking companies by their income \cite{pareto}.  Similarly, ranking US cities by their population or
citation analysis of scientific papers or many other distributions
yields the same law.  Recently, Watanabe \cite{watanabe} observed in numerical simulations that percolation clusters at the critical threshold also obey Zipf's Law, and Ma \cite{ma,ma05} proposed to make use of this finding to detect a crossing of the critical threshold.

One can derive \cite{BPA05} exact expressions for the average size of the
largest clusters in a fragment distribution
{\em at the critical point}, which is generated by a scaling
theory that obeys eq.~(\ref{eqa}).  Under the assumption
that the system size $V$ is large compared to the size of the emitted cluster, the
probabilities $p_k(i)$ to have $k$ clusters of size $A$ in a given event are Poissonian.
And the probability
that a cluster of size $A$ is the largest in a
given fragmentation event
\begin{equation}
	P_{1}(A) =  [1-p_0(A)]\cdot p_0(>\hspace*{-0.1cm}A)
	= [1-e^{-a\,A^{-\tau}}]\cdot e^{[a\, \zeta(\tau, 1+A) - a\, \zeta(\tau, 1+V)]},
\end{equation}
with the normalization constant $a = V / \sum_{A=1}^V A^{1-\tau}$. The probability
that a cluster is the second largest is
\begin{eqnarray}
	 P_{2} (A) &=& p_{ \ge 2} (A) \cdot p_0 ( > A) + p_{ \ge 1} (A) \cdot p_1 ( > A) \nonumber\\ 
  	&=& [1 - p_0 (A) - p_1 (A)] \cdot p_0 ( > A) + [1 - p_0 (A)] \cdot p_1 ( > A)
\end{eqnarray}
The probabilities for higher ranks follow from
recursion relations. From this we obtain the exact expression for the average size of the
largest clusters by summation over all $A$, $\langle A_{r} \rangle = \sum_{A=1}^{V} A\cdot P_{r}(A)$.

\begin{figure}[hbt]
\centerline{\includegraphics{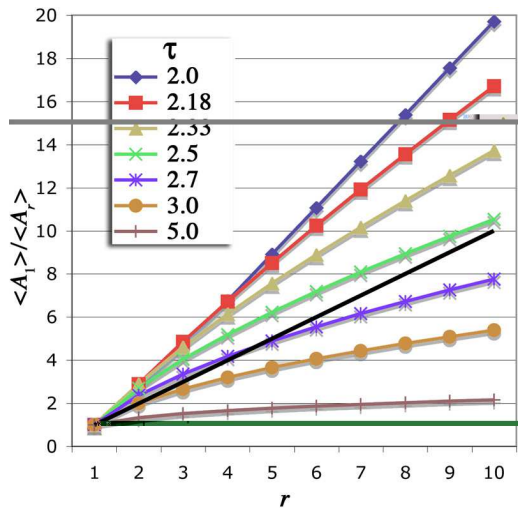}}
\caption{Average size of largest cluster over average size of cluster of rank $r$ as a function of $r$
for different values of the critical exponent $\tau$. The expectation from Zipf's law is represented
by the straight line without plot symbols.}
\label{fig3} 
\end{figure}

In fig.~\ref{fig3} we show the results (plot symbols)
from these (exact!) simulations for different values of
the critical exponent $\tau$ and for a system size of 10,000 constituents. We take the ratio of the
average size of the largest cluster to that of the cluster of rank $r$.  If Zipf's law were to hold, one
would expect a strictly linear function with slope one, if plotting this ratio versus the rank (straight line
without plot symbols).  One can see that different values of $\tau$ lead to different curves, and that
{\em none} strictly adheres to Zipf's law.

If one uses an expansion for our analytical results, one finds that the average cluster size as a function of rank $r$ follows a more general Zipf-Mandelbrot distribution \cite{mandel1,mandel2},
$\langle A_{r}\rangle = c (r + k)^{-\lambda}$, 
where the offset $k$ is an additional constant that one has to introduce, and $\lambda$ is asymptotically approximated as a function of the critical exponent $\tau$,
$\lambda(\tau) = 1/(\tau-1)$. These distributions are represented by the solid lines in fig.~\ref{fig3}.

Thus the result of our calculations is that the cluster size distributions do 
not follow Zipf's Law, but instead follow the more general Zipf-Mandelbrot distributions \cite{Cam05}.

This research was supported by the US National Science Foundation under
grants PHY-0245009 (W. Bauer) and PHY-0243709 (B. Alleman, Hope College, summer REU student
at MSU), as well as
the US Department of Energy under grant number DE-FG02-03ER41259 (S. Pratt).

%
\end{document}